\documentclass[floatfix,lengthcheck,showpacs,amssymb,amsmath,amsfonts,twocolumn,nofootinbib,longbibliography]{revtex4-1}
\usepackage{lineno}
\usepackage[utf8]{inputenc}
\usepackage{color}
\usepackage{graphicx}
\usepackage{float}
\usepackage{subfigure}
\usepackage{amsmath}
\usepackage{afterpage}
\usepackage{acronym}
\usepackage{multirow}
\usepackage{tikz}
\usetikzlibrary{arrows,shapes,trees,decorations.pathreplacing}
\usepackage{import}
\usepackage{svn-multi}
\usepackage{lineno}
\usepackage{hyperref}
\usepackage{gensymb}
\usepackage{longtable}
\usepackage{xcolor,colortbl}
\definecolor{Gray}{gray}{0.9}
\hypersetup{
    colorlinks=true,
    linkcolor=blue,
    filecolor=magenta,      
    urlcolor=cyan,
}

\newcommand{\msun}{\ensuremath{\mathrm{M}_{\odot}}}

\begin{document}


\title[]{Search for gravitational waves from the coalescence of sub-solar mass binaries in the first half of Advanced LIGO and Virgo's third observing run}

\author{Alexander H. Nitz}
\email{alex.nitz@aei.mpg.de}
\author{Yi-Fan Wang}
\affiliation{Max-Planck-Institut f{\"u}r Gravitationsphysik (Albert-Einstein-Institut), D-30167 Hannover, Germany}
\affiliation{Leibniz Universit{\"a}t Hannover, D-30167 Hannover, Germany}

\begin{abstract}
We present a search for gravitational waves from the coalescence of sub-solar mass black hole binaries using data from the first half of Advanced LIGO and Virgo's third observing run. The observation of a sub-solar mass black hole merger may be an indication of primordial origin; primordial black holes may contribute to the dark matter distribution. We search for black hole mergers where the primary mass is $0.1-7 M_{\odot}$ and the secondary mass is $0.1-1 M_{\odot}$.  A variety of models predict the production and coalescence of binaries containing primordial black holes; some involve dynamical assembly which may allow for residual eccentricity to be observed. For component masses $>0.5 M_{\odot}$, we also search for sources in eccentric orbits, measured at a reference gravitational-wave frequency of 10 Hz, up to $e_{10}\sim 0.3$. We find no convincing candidates and place new upper limits on the rate of primordial black hole mergers. The merger rate of 0.5-0.5 (1.0-1.0)~$M_{\odot}$ sources is $<7100~(1200)$ Gpc$^{-3}$yr$^{-1}$. Our limits are $\sim3-4$ times more constraining than prior analyses. Finally, we demonstrate how our limits can be used to constrain arbitrary models of the primordial black hole mass distribution and merger rate.
\end{abstract}

\date{\today}
\maketitle


\section{Introduction}

Gravitational-wave astronomy has entered an era of routine observations. The Advanced LIGO~\cite{TheLIGOScientific:2014jea} and Virgo~\cite{TheVirgo:2014hva} observatories have now completed three observing runs (O1, O2, and O3); each was accompanied by significant increases in sensitivity~\cite{Buikema:2020dlj}. To date, over 50 binary black hole mergers have been reported~\cite{Abbott:2020niy,Nitz:2021uxj, Venumadhav:2019lyq}. These observations have had significant impact on the study of the merger rate and population of compact objects~\cite{Abbott:2020gyp}; notable events confirm the likely existence of black holes with component masses in the pair-instability gap ($>50$ \msun) \cite{Abbott:2020tfl, Abbott:2020niy} or in the region 3-5 \msun \cite{Abbott:2020khf}. The possibility that these extremal parts of the distribution~\cite{Carr:2019kxo,Clesse:2020ghq,Vattis:2020iuz,DeLuca:2020sae} or a fraction of the bulk of observed mergers~\cite{Franciolini:2021tla,DeLuca:2021wjr,Jedamzik:2020omx,Garcia-Bellido:2020pwq,Bird:2016,Clesse:2016vqa,Sasaki:2016jop,DeLuca:2020qqa,Hutsi:2020sol} may be due to the coalescence of primordial black holes (PBHs) is under active investigation.

Currently, there is no clear observational evidence for the existence of PBHs. However, in addition to providing an explanation for some of the observed LIGO and Virgo mergers~\cite{Clesse:2020ghq}, primordial black holes may be the origin of some observed microlensing incidents~\cite{Niikura:2019kqi}, excess cross-correlation between cosmic x-ray and cosmic microwave background ~\cite{Hasinger:2020ptw}, the current excess in gravitational-wave background observed by NANOGrav~\cite{PhysRevLett.126.041303}, and the seeds for galaxy and supermassive black hole formation~\cite{Rubin:2001yw,Khlopov:2002yi,Khlopov:2004sc}. 
Many of these observations are also consistent with more mundane explanations and standard stellar-formation scenarios~\cite{Abbott:2020gyp}. In contrast, there are no known mechanisms through standard stellar evolution to produce sub-solar mass black holes; the observation of a single sub-solar mass black hole would be decisive for the existence of primordial black holes or for even more exotic scenarios such as dark matter triggered formation of black holes~\cite{Shandera:2018xkn,Singh:2020wiq,Dasgupta:2020mqg}.

\begin{figure}[tb!]
    \centering
    \includegraphics[width=\columnwidth]{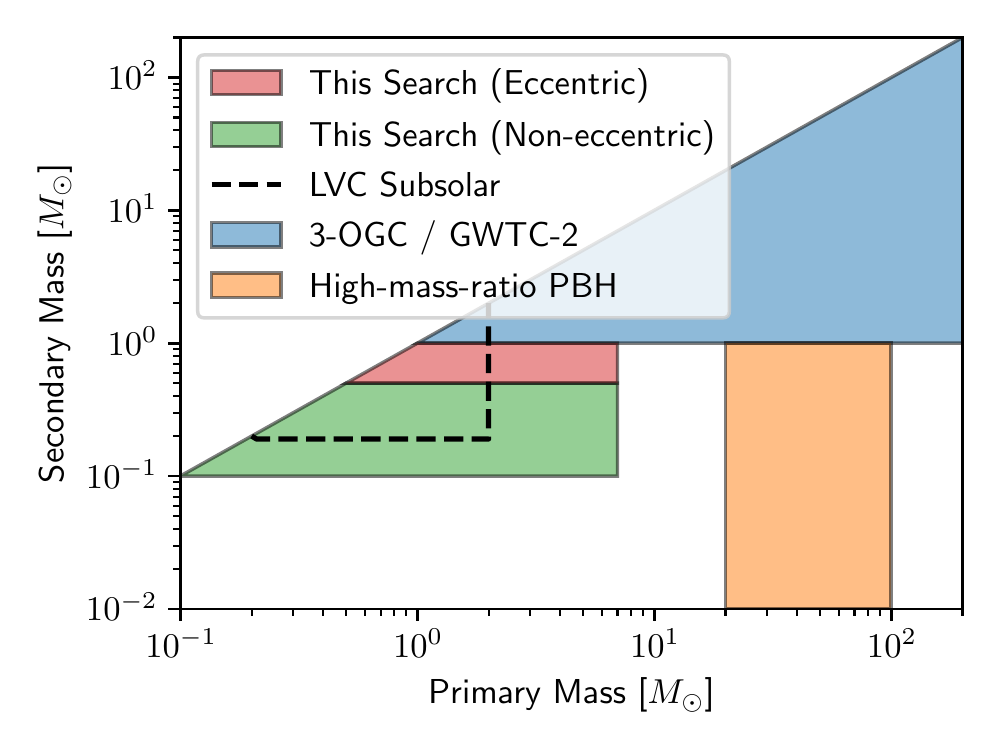}
    \caption{The regions searched by recent gravitational-wave analyses of the LIGO and Virgo data as a function of detector-frame primary and secondary mass. The region we search for non-eccentric sources (green) and the region sensitive to sources with eccentricity up to $e_{10}\sim 0.3$ (red) are shown. For comparison, we include the search region of the most recent sub-solar mass search by the LVC~\cite{Authors:2019qbw} (dashed), the search for high-mass-ratio mergers~\cite{Nitz:2020bdb}, and searches for standard stellar-mass sources~\cite{Nitz:2021uxj,Abbott:2020niy}. The axes are truncated at 200~$M_{\odot}$. 
    }
    \label{fig:searches}
\end{figure}

Several searches for gravitational waves from the coalescence of sub-solar mass mergers have already been conducted using data from LIGO's first two observing runs (O1, O2); these include searches for comparable mass binary black holes~\cite{Abbott:2018oah,Authors:2019qbw,Phukon:2021cus}, eccentric mergers~\cite{Nitz:2021mzz}, and high-mass-ratio sources~\cite{Nitz:2020bdb}. No likely candidates have been found. In this paper we report a search for gravitational waves from the coalescence of black holes with primary mass $0.1-7\msun$ and secondary mass $0.1-1\msun$ using the open data from the first half of the third observing (O3a) run of Advanced LIGO and Virgo. The parameter space for our search region is shown in Fig.~\ref{fig:searches}. 
The most significant candidate in our search has a false alarm rate of 1 per O(month). Given the time observed, we consider our results consistent with a null observation and place new limits on the rate of sub-solar mass mergers which are $3-4$ times more stringent than prior analyses due to the significant improvement of sensitivity and detector robustness of O3 compared to O1 and O2 \cite{Abbott:2020niy}.

Our limits on the rate of sub-solar mass mergers can be related to constraints on the fraction of dark matter composed of PBHs; this requires a model of the binary's formation to predict the PBH abundance from the observed merger rate. Current astrophysical models have large uncertainties in their predictions of both the black hole mass function and binary formation rate~\cite{Nakamura:1997sm,Sasaki:2016jop,Bird:2016,10.1103/PhysRevD.99.043533,Clesse:2016vqa,Chen:2018czv,Ali-Haimoud:2017rtz}. Primordial black holes may form binaries in the early universe if they can decouple from the cosmic expansion. However, it is under investigation what fraction of binaries would be disrupted in the following evolution. Ref.~\cite{Raidal:2018bbj} shows with N-body simulations that a significant fraction would be disrupted if the fraction that primordial black hole contribute to the dark matter is $f_\mathrm{PBH}=100\%$. Primordial black holes can also form binaries in the late universe by dynamical capture due to gravitational-wave dissipation~\cite{Bird:2016,Fakhry:2020plg,Fakhry:2021tzk}; this scenario may also lead to residual eccentricity by the time it is observable by gravitational-wave detectors~\cite{Wang:2021qsu}. However, the event rate of binaries formed in the early universe is expected to be dominant compared with the late universe channel, depending on the intensity of binary disruption~\cite{Raidal:2018bbj}.

Due to the wide variety of models, in this paper, we consider a fiducial model which assumes a monochromatic distribution of primordial black hole mass. The same model was used in past sub-solar mass searches~\cite{Abbott:2018oah,Authors:2019qbw,Phukon:2021cus,Nitz:2020bdb,Nitz:2021mzz}, and we include for comparison purposes. We also consider the uncertainty on the rate estimates arising from the fraction of binaries that are disrupted after formation~\cite{Raidal:2018bbj}. Constraints for specific models with broad mass distributions can be derived from our observational constraints.

\section{Search}

We conduct our analysis in a similar manner to our previously presented search of the first two observing runs~\cite{Nitz:2021mzz}; however, for the first time we include data from the Virgo observatory. We use the open-source PyCBC toolkit~\cite{Usman:2015kfa, pycbc-github} to conduct a matched-filtering based search~\cite{Allen:2005fk}. Matched filtering allows us to extract a potential signal using the predicted gravitational waveform as a template. Potential candidates are assessed for consistency between the operating observatories~\cite{Nitz:2017svb} and against the expected morphology of the gravitational waveform~\cite{Nitz:2017lco,Allen:2004gu}. Each candidate is assigned a ranking statistic value that takes into account these factors in addition to the measured noise variance~\cite{Davies:2020,Mozzon:2020gwa}.

The statistical significance of each candidate is assessed by comparing to an empirically measured distribution of false alarms~\cite{Usman:2015kfa}. The distribution is measured by conducting numerous fictitious analyses whereby we offset detectors' data in time. This procedure purposefully violate the time-of-flight constraints between the detectors to remove coincident astrophysical sources and create analyses containing only false alarms~\cite{Was:2009vh,2017PhRvD..96h2002C,TheLIGOScientific:2016zmo}.

As matched filtering requires accurate models of the expected gravitational-wave signal, we use a combination of the TaylorF2 post-Newtonian approximant accurate to 3.5PN~\cite{Sathyaprakash:1991mt,Droz:1999qx,Blanchet:2002av,Faye:2012we} and the TaylorF2e model~\cite{Moore:2019a,Moore:2019b,Moore:2018}. TaylorF2e is an extension of TaylorF2 which includes corrections for moderate eccentricity. Both TaylorF2 and TayloF2e model only the inspiral portion of a gravitational-wave signal and do not account for the phase where the binary finally merges. The merger can be safely neglected as we search for sources only up to a total mass of $8~\msun$. For these sources, the merger occurs at a frequency above the most sensitive band of the instruments. 

To search for a broad region, we use the stochastic algorithm~\citep{Harry:2009ea} to create a discrete bank of templates  designed to ensure that we recover $>95\%$ of a signal's signal-to-noise ratio if it has parameters within the boundaries of our search. Our bank is designed to recovery binaries in quasi-circular orbits where the primary mass is $0.1-7~\msun$ and the secondary mass is $0.1-1~\msun$. In addition, for sources with component masses $>0.5 \msun$, the bank is designed to recover sources with eccentric orbits up to $e_{10} \sim 0.3$, where $e_{10}$ is the eccentricity at the fiducial dominant-mode gravitational-wave frequency of 10 Hz. 
We assume that PBHs will have negligible spin; this is consistent with the predictions of PBH spin distributions~\cite{Chiba:2017rvs,DeLuca:2019buf,DeLuca:2020bjf,Mirbabayi:2019uph,Postnov:2019pkd}. To save on computational cost, we limit the starting frequency of each template so that its duration is $<512s$; otherwise a cutoff at a gravitational-wave frequency of 20 Hz is used. The template bank was also constructed with this lower frequency criteria.
These choices result in a bank with $\sim 7.8$ million templates, where $50\%$ of the templates have nonzero eccentricity and use the TaylorF2e model.

\section{Observational Results}

We search for gravitational waves from the coalescence of sub-solar mass compact binaries using the public LIGO and Virgo data from the first half of the third observing run (O3a)~\cite{Vallisneri:2014vxa,Abbott:2019ebz}; data from the second half of the observing run is not yet available. We analyze the nearly 150 days of data where at least two observatories were operating; the twin LIGO observatories were operating for $\sim 100$ days of this period. In comparison to the previous observing run, the LIGO instruments had $\sim30-50\%$ greater range~\cite{Buikema:2020dlj}.

The most significant candidates from our search are identified at a false alarm rate of one per O(month) and shown in Table~\ref{table:searchresults}; this is consistent with a null observation and our expectation that the noise candidates follow a Poisson distribution. Under the assumption of a null detection, we place limits on the rate of binary mergers at $90\%$ confidence using the loudest event method of Ref.~\cite{Biswas:2007ni}. The limit on the merger rate $R_{90}$ is given as
\begin{equation}
    R_{90} = \frac{2.3}{VT}
    \label{eq:singlerate}
\end{equation}
where V is the estimated sensitive volume of the analysis assessed at the false alarm rate of the most significant observed candidate and $T$ is the duration of the observation period. We estimate the surveyed volume-time of our analysis by measuring our analysis' response to a simulated population of $O(10^5)$ sources. We assume that sources are isotropically distributed in their orientations and sky location in addition to a uniform distribution in volume. We measure the mass dependence of our sensitive volume using separate simulation sets each with fixed source masses. For simulations that include eccentricity, we assume a uniform distribution where $e_{10} \in [0, 0.3)$.

\setlength{\tabcolsep}{2mm}
\begin{table}
  \caption{The top 5 candidates in our search with the highest inverse false alarm rates (IFAR). The GPS time of each candidate, the component mass $m_{1/2}$ and the eccentricity of the template chosen by the search are shown.}
  \label{table:searchresults}
\begin{tabular}{ccccc}
   GPS time 
  & IFAR (yr) 
  & $m_1 / M_\odot$ 
  & $m_2 / M_\odot$ 
  &$e_\textrm{10}$  \\\hline

    \hline
1245411568.354 & 0.084 & 0.69 & 0.21 & 0.00  \\
  1242817372.434 & 0.079 & 0.86 & 0.11 & 0.00 \\
1246418221.718 & 0.075 & 0.13 & 0.13 & 0.00 \\
1252963276.322 & 0.062 & 1.05 & 0.52 & 0.28 \\
1240000657.632 & 0.057 & 3.04 & 0.10 & 0.00 \\
\end{tabular}
\end{table}

\begin{figure}[tb!]
    \centering
    \includegraphics[width=\columnwidth]{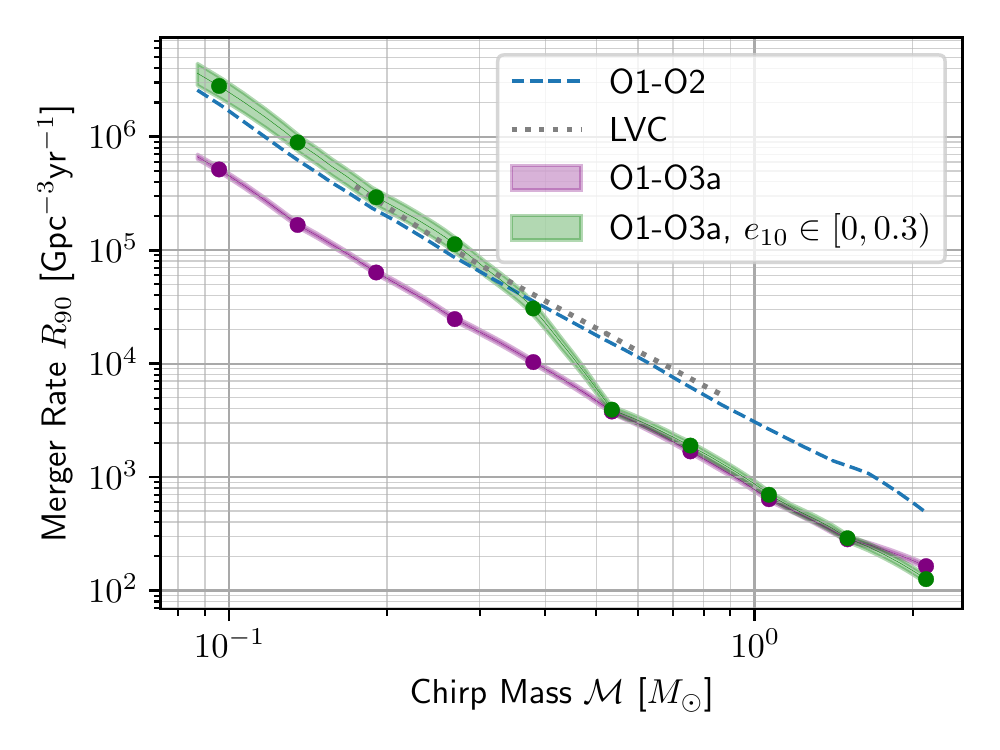}
    \caption{Upper limit on the rate of mergers at 90\% confidence ($R_{90}$) for our search (purple). For comparison, we show the previous limits from our search of the O1 and O2 data (blue dashed)~\cite{Nitz:2021mzz} in addition to the most recent results from sub-solar masses searches conducted by the LVC of the O2 data (black dotted)~\cite{Authors:2019qbw}. These all assume that sources are quasi-circular; constraints assuming a uniform distribution of $e_{10} \in [0, 0.3)$ are also shown (green). As expected, the limit for eccentric sources closely matches that for quasi-circular binaries where the component masses are $>0.5 \msun$. Shaded regions show the one sigma uncertainty on the rate due to the monte-carlo estimation of the search's surveyed volume-time.}
    \label{fig:rate}
\end{figure}

The resulting upper limit on the merger rate, combined with the limit from the prior analysis of O1 and O2~\cite{Nitz:2021mzz}, is shown as a function of chirp mass in Fig.~\ref{fig:rate}, where chirp mass is defined as $\mathcal{M} = (m_1m_2)^{3/5}/(m_1+m_2)^{1/5}$ and $m_{1,2}$ are the masses of the two components of a binary. For sources within our target parameter space, this limit also holds for sources with varied mass ratios, but the same chirp mass; a similar conclusion was noted in Ref.~\cite{Authors:2019qbw}. Our results limit the merger rate for 0.1-0.1, 0.5-0.5, and 1.0-1.0~\msun{} sources to $<670000$, 7100, and 1200 Gpc$^{-3}$yr$^{-1}$, respectively. This is an improvement between $3-4$ times over previous analyses which only used data from the first two observing runs.

\section{Implications for primordial black hole abundance}

We use the observational upper limits on the sub-solar mass compact binary merger rate to constrain models of primordial black hole binary formation. Existing models have significant uncertainties on the primordial black hole mass distribution and the binary formation rate \cite{Nakamura:1997sm,Sasaki:2016jop,Bird:2016,10.1103/PhysRevD.99.043533,Clesse:2016vqa,Chen:2018czv,Ali-Haimoud:2017rtz,Carr:2019kxo}.
Constraints on models can be derived from the observational event rate upper limits given a specific primordial black hole mass distribution.
We consider a fiducial mass distribution, a delta distribution following Refs.~\cite{Abbott:2018oah,Authors:2019qbw} and our previous work \cite{Nitz:2020bdb,Nitz:2021mzz} to allow for a consistent comparison. 

For the binary formation rate, we consider the mechanism initially proposed by \cite{Nakamura:1997sm} and developed by \cite{Sasaki:2016jop,Chen:2018czv,Ali-Haimoud:2017rtz,Raidal:2018bbj,Vaskonen:2019jpv,Hutsi:2020sol} where primordial black holes form bounded binaries in the early universe and merge recently. 
The binaries form in the late universe is subdominant \cite{Wang:2021qsu}, therefore we use the search results from circular binaries to constrain the early universe formation model. 
The binary merger rate is given by Refs.~\cite{Chen:2018czv,Ali-Haimoud:2017rtz} for a general mass distribution $P(m)$
\begin{eqnarray}\label{eq:broadmassrate}
&R&(\widetilde{f}_\mathrm{PBH},m_1,m_2) = 3\times10^6 \widetilde{f}_\mathrm{PBH}^2(0.7\widetilde{f}_\mathrm{PBH}^2 + \sigma_\mathrm{eq}^2)^{-\frac{21}{74}} \nonumber \\
&\times&(m_1m_2)^{\frac{3}{37}}(m_1+m_2)^{\frac{36}{37}}\mathrm{min}\left(\frac{P(m_1)}{m_1},\frac{P(m_2)}{m_2}\right
) \nonumber \\
&\times&\left(\frac{P(m_1)}{m_1}+\frac{P(m_2)}{m_2}\right) ~\mathrm{Gpc}^{-3} \mathrm{yr}^{-1},
\end{eqnarray}
where $R\cdot dm_1dm_2$ is the event rate at the binary component masses $m_{1/2}~ \msun{}$.
The normalization for mass distribution is $\int P(m)dm=1$.
The parameter $\sigma_{eq}=0.005$ is the variance of dark matter density perturbation at the matter radiation equality epoch \cite{Ali-Haimoud:2017rtz}, and we keep this factor separate from other binary disruption effects to perform consistent comparison with previous work.
Ref.~\cite{Raidal:2018bbj} has used an N-body simulation to show most binaries after formation would be disrupted by their environment if $f_\mathrm{PBH}=100\%$, and thus introduced a suppression factor $S$ with value $<1$ which accounts for the disruption as a function of primordial black hole mass and fraction. 
For a relatively narrow mass distribution, Ref.~\cite{Hutsi:2020sol} shows $S$ (referred to as $S_2$ in Ref.~\cite{Hutsi:2020sol}) is estimated to be $1\%$ for $f_\mathrm{PBH}=100\%$ and $\sim 1$ for $f_\mathrm{PBH}<1\%$ and the suppression is only a function of $f_\mathrm{PBH}$.
However, underestimation by one order of magnitude may exist for $f_\mathrm{PBH}=100\%$ if the merger of binaries perturbed by the environment after formation is included~\cite{Vaskonen:2019jpv,Hutsi:2020sol,Phukon:2021cus}.
To account for this uncertainty, we define an effective fraction which relates to the true fraction by $\widetilde{f}_\mathrm{PBH}^{53/37} = S f_\mathrm{PBH}^{53/37}$.
Constraints in our past work \cite{Nitz:2020bdb,Nitz:2021mzz} on $f_\mathrm{PBH}$ should also be understood as constraints on $\widetilde{f}_\mathrm{PBH}$; they implicitly assume the suppression factor is unity.
The true fraction $f_\mathrm{PBH}$ can be recovered with a well-understood estimation of $S$.

\begin{figure}[tb!]
    \centering
    \includegraphics[width=\columnwidth]{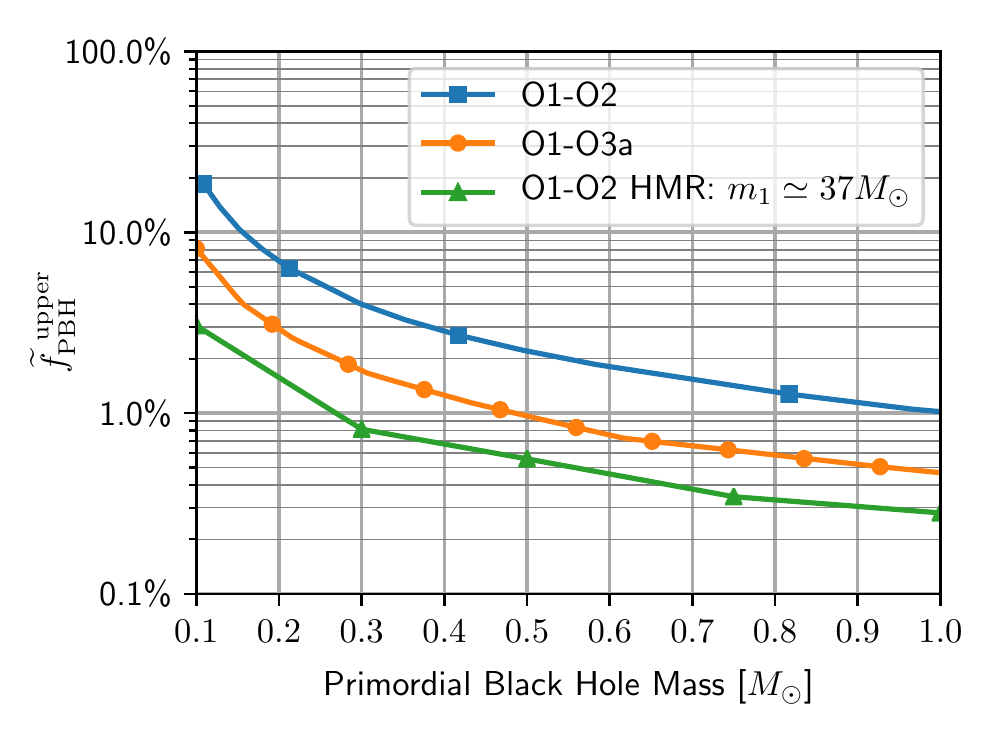}
    \caption{The upper limits on the effective fraction of the primordial black hole contribution to dark matter, $\widetilde{f}_\mathrm{PBH}$, from the search for sub-solar mass compact binaries in LVC O1-O3a data.
    As a comparison, we also plot the constraints from O1-O2 data \cite{Nitz:2021mzz} for the monochromatic mass distribution and for the two-point mass distribution of high-mass-ratio binaries~\cite{Nitz:2020bdb} where the primary mass is fixed to the average of the 2-OGC events $\simeq 37\msun$. Note that this latter limit requires the additional assumption that the primary mass abundance be consistent with accounting for the majority of LIGO-observed BBH mergers.
    }
    \label{fig:pbhfraction}
\end{figure}

For a delta distribution of mass, Eq.~\ref{eq:pbhrate} is reduced to
\begin{eqnarray}\label{eq:pbhrate}
R(\widetilde{f}_\mathrm{PBH},m) =3\cdot10^6 \widetilde{f}_\mathrm{PBH}^2(0.7\widetilde{f}_\mathrm{PBH}^2 + \sigma_\mathrm{eq}^2)^{-\frac{21}{74}}m^{-\frac{32}{37}}.
\end{eqnarray}
Using the observational rate limit given in Fig.~\ref{fig:rate}, the upper limit on the effective fraction of primordial black hole with a delta distribution of mass is shown in Fig.~\ref{fig:pbhfraction}. 
A comparison with previous search results for sub-solar mass compact binaries is also plotted.
The results should be interpreted as constraints on the combined effect on the binary merger rate from the abundance of primordial black holes and the merger rate suppression due to environmental interaction.
Future improvements on theoretical modeling and observational results can resolve the entanglement.
Overall, results in Fig.~\ref{fig:pbhfraction} show that our constraints are $\sim 2-3$ times tighter than \cite{Nitz:2021mzz} using O1 and O2 data.

Constraints can also be derived on primordial black hole scenarios which predict broad mass distributions. By extending Eq.~\ref{eq:singlerate}, the corresponding $90\%$ upper limits can be obtained by requiring
\begin{equation}\label{eq:general}
    \int R(\vec{\theta},m_1,m_2)VT(m_1, m_2) dm_1dm_2 = 2.3,
\end{equation}
where R is the model-predicted merger rate density as a function of the component masses and may include additional model parameters $\vec{\theta}$. VT is the surveyed volume-time of our analysis as a function of the component masses; this is available as part of our data release.

To illustrate, we consider the scenario that primordial black holes are produced in the early Universe QCD phase transition era \cite{Carr:2019kxo}. This model is shown to have a peak at $\sim1\msun{}$ and to be able to account for the origin of the extremal parts of the already observed binary black hole merger distribution.
We choose the mass distribution $P(m)$ by following Ref.~\cite{Carr:2019kxo} from the early universe scalar perturbation spectral index 0.96 and use the binary formation rate of Eq.~\ref{eq:broadmassrate}. 
We assume a mass-independent suppression factor may be applicable here given the narrow peak of $P(m)$.
The observational upper limits on event rate from our search require $\widetilde{f}_\mathrm{PBH}\leq1\%$ in this scenario. For more general mass distributions, Eq.~\ref{eq:general} is capable of constraining the PBH abundance or other model parameters given a specific primordial black hole mass distribution and prescription which can predict the resulting merger rate density $R(\vec{\theta},m_1,m_2)$.

\section{Conclusions}
We conduct a search for gravitational-waves from the coalescence of sub-solar mass black holes using data from the first half of the third observing run of Advanced LIGO and Virgo. We find no clear detections and so place new limits on the rate of mergers. The increased sensitivity of the O3a data allows us to improve upon the state-of-the-art limits by $3-4$ times. The second half of the O3 data (O3b) would be expected to improve these limits by another factor of $1.5-2$ times.

We apply our rate limits to a fiducial monochromatic mass distribution and compare our results to the limits from prior analyses. Overall the constraints on the effective fraction of primordial black holes in dark matter are 2-3 tighter than the previous results. 
Our observational results can also help resolve the event rate modeling uncertainties by constraining model parameters. 
If we assume that dark matter consisted entirely of black holes, we constrain the suppression factor to $S\leq 0.1\%$ for 0.5-0.5~\msun{} mergers.

Lastly, we demonstrate how to apply our limits to models that can predict the rate of PBH mergers.
To aid in comparing our rate limits to the various formation scenarios leading to sub-solar mass mergers, we make the detailed constraints available at \url{https://github.com/gwastro/subsolar-o3a-search}. In addition, we make available the configuration files and template bank necessary to reproduce the analysis.

\vspace{54pt}
\acknowledgments

 We acknowledge the Max Planck Gesellschaft. We thank the computing team from AEI Hannover for their significant technical support. This research has made use of data from the Gravitational Wave Open Science Center (https://www.gw-openscience.org), a service of LIGO Laboratory, the LIGO Scientific Collaboration and the Virgo Collaboration. LIGO is funded by the U.S. National Science Foundation. Virgo is funded by the French Centre National de Recherche Scientifique (CNRS), the Italian Istituto Nazionale della Fisica Nucleare (INFN) and the Dutch Nikhef, with contributions by Polish and Hungarian institutes.

\bibliography{references}

\end{document}